\documentclass[prd,superscriptaddress,footinbib,twocolumn,floatfix]{revtex4}
\usepackage{amsmath,amssymb,amsfonts,bm,amscd}
\usepackage{graphics}
\usepackage{natbib,hyperref}
\usepackage{xfrac}

\makeatletter
\renewcommand{\@makecaption}[2]{
  \vskip\abovecaptionskip
  \sbox\@tempboxa{\small\sf #1: #2}%
  \ifdim \wd\@tempboxa >\hsize
  \small\sf #1: #2\par
  \else
    \global \@minipagefalse
    \hb@xt@\hsize{\hfil\box\@tempboxa\hfil}%
  \fi
  \vskip\belowcaptionskip}
\makeatother

\def\ba{\begin{eqnarray}}
\def\ea{\end{eqnarray}}

\def\tilde{\widetilde}

\def\bar{\overline}

\def\Dslash{\,\,{\raise.15ex\hbox{/}\mkern-12mu D}}
\def\Dbarslash{\,\,{\raise.15ex\hbox{/}\mkern-12mu {\bar D}}}
\def\delslash{\,\,{\raise.15ex\hbox{/}\mkern-9mu \partial}}
\def\delbarslash{\,\,{\raise.15ex\hbox{/}\mkern-9mu {\bar\partial}}}
\def\pslash{\,\,{\raise.15ex\hbox{/}\mkern-9mu p}}
\def\calDslash{\,\,{\raise.15ex\hbox{/}\mkern-12mu {\cal D}}}

\renewcommand{\bar}{\overline}

\begin{document}

\title{Rationality in Four Dimensions}

\author{Leonardo Rastelli}

\affiliation{C.\ N.\ Yang Institute for Theoretical Physics, Stony Brook University, Stony Brook, NY 11794}

\author{Brandon C.\ Rayhaun}
\affiliation{C.\ N.\ Yang Institute for Theoretical Physics, Stony Brook University, Stony Brook, NY 11794}

\begin{abstract}
By leveraging the physics of the Higgs branch, we argue that the conformal central charges $a$ and $c$ of an arbitrary 4d $\mathcal{N}=2$ superconformal field theory (SCFT) are rational numbers. Our proof of the rationality of $c$ is conditioned on a well-supported conjecture about how the Higgs branch of an SCFT is encoded in its protected chiral algebra. To establish the rationality of $a$, we further rely on a widely-believed technical assumption on the high-temperature limit of the superconformal index. 
\end{abstract}

\maketitle

\subsection*{Introduction}

Four-dimensional conformal field theories (CFTs) with eight real supercharges possess rich sectors of observables protected by supersymmetry. These sectors often organize themselves into tightly constrained but highly non-trivial mathematical structures which render many aspects of their physics tractable. A prime example is that the algebra of Schur operators is encoded in an auxiliary two-dimensional chiral algebra, also known as a vertex operator algebra (VOA) \cite{Beem:2013sza}. The relative ease with which one may carry out various computations within these protected sectors has lead to an abundance of data pertaining to 4d $\mathcal{N}=2$ SCFTs, which has correspondingly inspired many conjectures concerning their properties. It is certainly of interest to produce physical arguments for such conjectures, or to at least  draw the logical arrows between them in the cases that full proofs are beyond reach.

In this letter, we study the most basic invariants of a 4d CFT: its central charges. In particular, we consider the $a$ and $c$ trace anomalies which, as the name suggests, can be defined through the trace of the stress tensor,
\begin{align}
    \langle T_{\mu}^\mu\rangle = \frac{1}{16\pi^2}(c W^2 -a E)
\end{align}
where here, $W$ and $E$ are the Weyl and Euler curvature invariants, respectively, of the spacetime manifold, 
\begin{align}
\begin{split}
    W^2 &= R_{\mu\nu \rho\sigma} R^{\mu\nu\rho\sigma}-2R_{\mu\nu}R^{\mu\nu}+\frac13 R^2 \\ 
    E &= R_{\mu\nu\rho\sigma}R^{\mu\nu\rho\sigma}-4R_{\mu\nu}R^{\mu\nu}+R^2.
\end{split}
\end{align}
If the theory has a flavor symmetry with a non-Abelian simple factor $\mathfrak{g}$, we may also study the corresponding flavor central charge $k$, which is diagnosed through the leading behavior of the two-point function of the conserved currents as 
\begin{align}\label{eqn:flavorcc}
    \langle J^A_\mu(x) J^B_\nu(0)\rangle = \frac{3k}{4\pi^4}\delta^{AB} \frac{x^2g_{\mu\nu}-2x_\mu x_\nu}{x^8}+\cdots
\end{align}
where $A$ and $B$ are adjoint indices for $\mathfrak{g}$.
The central charges $a$, $c$, and $k$ of a CFT play starring roles in various physical applications. For example, $a$ famously decreases under renormalization group flow \cite{Komargodski:2011vj}. When we want to emphasize the four-dimensional nature of these quantities, we will write them as $a_{4\mathrm{d}}$, $c_{4\mathrm{d}}$, and $k_{4\mathrm{d}}$.

In CFTs with at least $\mathcal{N}=1$ supersymmetry, it is often possible to determine the anomalies exactly via a variety of methods \cite{Anselmi:1997am,Anselmi:1997ys,Intriligator:2003jj,Kutasov:2003iy,Argyres:2007cn,Aharony:2007dj,Shapere:2008zf,Martone:2020nsy}. 
Computations in large classes of examples support the lore that all  $\mathcal{N}=2$ SCFTs have rational central charges, while  in $\mathcal{N}=1$ SCFTs they are generally irrational (though conjecturally algebraic~\cite{Intriligator:2003jj}). 
In addition to its general intrigue, this lore, if true, has practical applications: For example, it furnishes a necessary condition for a 4d $\mathcal{N}=1$ Lagrangian to have enhanced $\mathcal{N}=2$ superconformal symmetry in the IR, a phenomenon which has been put to good use \cite{Maruyoshi:2016tqk,Maruyoshi:2016aim,Agarwal:2016pjo} in determining the superconformal indices of strongly-interacting isolated SCFTs like Argyres--Douglas theories \cite{Argyres:1995jj}. We henceforth refer to the assertion that $a$, $c$, and $k$ are rational in 4d $\mathcal{N}=2$ SCFTs as the \emph{Rationality Conjecture}.

The plausibility of the Rationality Conjecture is bolstered by a number of insights which have come from the beautiful program of leveraging the geometry of the Coulomb branch to classify 4d $\mathcal{N}=2$ SCFTs (see e.g.~\cite{Argyres:2015ffa,Argyres:2015gha,Argyres:2016xmc,Argyres:2016xua,Martone:2020nsy,Argyres:2020wmq,Martone:2021ixp}, and \cite{Martone:2020hvy} for a recent review). In particular, by applying anomaly matching arguments to the low energy effective action on the Coulomb branch, it is possible to derive explicit formulae for the  central charges $a$, $c$, and $k$ in a large class of theories (see equations 1.1a--1.1c of \cite{Martone:2020nsy}). For example, a celebrated formula of Shapere and Tachikawa~\cite{Shapere:2008zf} gives the relation
\begin{align}\label{eqn:ShapereTachikawa}
    2a-c = \frac14\sum_{i=1}^{\bf r}(2\Delta_{\mathcal{O}_i}-1)\, ,
\end{align}
where $\bf r$ is the rank of the theory, and the $\mathcal{O}_i$ are the generators of the Coulomb chiral ring, whose vacuum expectation values parametrize the Coulomb branch. When supplemented with the rationality of the scaling dimensions $\Delta_{\mathcal{O}_i}$ \cite{Argyres:2018urp,Caorsi:2018zsq}, Equation \eqref{eqn:ShapereTachikawa} implies the rationality of $2a-c$, and it turns out that the more general formulae of \cite{Martone:2020nsy} are sufficient to establish the rationality of $a$, $c$ and $k$ separately. (These arguments are unavailable when the Coulomb branch is empty, but in these cases one may appeal to the belief that all rank-zero SCFTs are theories of free hypermultiplets and discrete gaugings thereof.) 

One shortcoming of the Coulomb branch formulae is that they
rely on several geometric assumptions and physical arguments whose range of validity is not always completely clear.
For example, the standard analysis assumes that the Coulomb chiral ring is freely generated, and
 easy counterexamples to the Coulomb branch formulae
 can be obtained lifting this assumption by gauging discrete symmetries \cite{Argyres:2016yzz}. 
 Although such theories clearly still obey the Rationality Conjecture,
it is not obvious that \emph{all} theories which violate the Coulomb branch formulae must take this form. For this reason, we consider it worthwhile to revisit the problem using alternative methods. 

In this spirit, we offer a complementary approach to the Rationality Conjecture as it pertains to $a$ and $c$, which is based on the analysis of observables, like the Schur limit of the superconformal index, which are sensitive to the physics of the Higgs branch. A Higgs branch approach to the rationality of flavor central charges $k$ is largely left to future work, though we offer a few parting thoughts in this direction at the end of the paper. Throughout our analysis, we clearly articulate our assumptions, which we believe are comparatively conservative. A pleasant byproduct of our investigation is that, when suitably interpreted as statements about VOAs, our arguments become mathematically rigorous. For example, we are able to prove that, for any quasi-lisse vertex operator algebra (with additional standard technical conditions assumed), the conformal central charge and the holomorphic scaling dimensions of ordinary simple modules are all rational. We begin by reviewing the relationship between 4d $\mathcal{N}=2$ SCFTs and chiral algebras. 

\subsection*{Review of the Protected Chiral Algebra}

In 4d SCFTs, operators in short representations are counted (up to equivalence relations that account for the possible recombinations of short multiplets into long ones) by the superconformal index \cite{Kinney:2005ej}. The  superconformal index can equivalently be defined as the partition function of the theory on a continuous family of backgrounds with $S^3\times S^1$ topology. 
The  index admits many interesting special limits. The one  useful for our purposes is the Schur limit~\cite{Gadde:2011ik,Gadde:2011uv}, 
\begin{align}
    \mathcal{I}_{\mathrm{Schur}}(q)=\mathrm{Tr}(-1)^F q^{E-R},
\end{align}
where  $E$ is the generator of dilatations and $R$ is the generator of the Cartan of the $\mathfrak{su}(2)_R$ symmetry. 
In this limit, the index only receives contributions from Schur operators which, by definition, satisfy the shortening conditions 
\begin{align}
\begin{split}
E- (j_1 +j_2)-2R&=0\\
    j_1 -j_2 -r&=0
\end{split}
\end{align}
%
where $r$ is the generator of $\mathfrak{u}(1)_r$, and $j_1$ and $j_2$ are the Cartan generators of the $\mathfrak{su}(2)_1\times \mathfrak{su}(2)_2$ isometries of the spatial $S^3$. The expectation values of Schur operators with $j_1=j_2=r=0$ parametrize the Higgs branch moduli space, and such operators are correspondingly referred to as Higgs branch operators. 

The vector space of Schur operators  
is endowed with a surprising additional structure~\cite{Beem:2013sza}. 
``Twisted-translated'' Schur operators (on a two-dimensional plane away from the origin) 
reside in the cohomology of a certain nilpotent supercharge,
 and their operator algebra within this cohomology furnishes the structure of a two-dimensional chiral algebra/VOA $\mathbb{V}(\mathcal{T})$.  
The chiral algebra $\mathbb{V}(\mathcal{T})$ is half-integer graded by the conformal dimensions $h$ of its operators,
\begin{align}
    \mathbb{V}(\mathcal{T}) = \bigoplus_{h \in \frac12 \mathbb{Z}} \mathbb{V}(\mathcal{T})_h
\end{align}
which are related to the quantum numbers in 4d as
\begin{align}\label{eqn:conformaldim}
    h=\frac12(E+j_1+j_2).
\end{align}
The central charge $c_{2\mathrm{d}}$ of  $\mathbb{V}(\mathcal{T})$ is related to the central charge $c_{4\mathrm{d}}$ of the SCFT $\mathcal{T}$ as 
\begin{align}
    c_{2\mathrm{d}} = -12 c_{\mathrm{4d}}.
\end{align}
If $\mathcal{T}$ possesses a flavor symmetry which has a non-Abelian simple factor $\mathfrak{g}$ with central charge $k_{4\mathrm{d}}$, the VOA contains an affine Kac--Moody subalgebra $(\mathfrak{g})_{k_{2\mathrm{d}}}\subset \mathbb{V}(\mathcal{T})$ with 
\begin{align}
    k_{2\mathrm{d}}=-\frac12 k_{4\mathrm{d}}.
\end{align}
This VOA ``categorifies'' the Schur index in the sense that its vacuum character recovers $\mathcal{I}_{\mathrm{Schur}}(q)$, up to a factor of $q^{-c_{2\mathrm{d}}/24}$ coming from the Casimir energy of the chiral algebra.

The protected chiral algebra of a 4d $\mathcal{N}=2$ SCFT satisfies a number of conditions that are typically assumed in mathematical treatments of vertex operator algebras. 
For example, it follows from Equation \eqref{eqn:conformaldim} that $h\geq0$, and from the uniqueness of the vacuum of $\mathcal{T}$ that $\mathbb{V}(\mathcal{T})_0  \cong \mathbb{C}$. As is standard, we also assume that the 4d spectrum is compact so that $\dim\mathbb{V}(\mathcal{T})_h<\infty$. A VOA which satisfies these three properties together is said to be of CFT-type.  
We also always assume that we have quotiented out by any null vectors in the chiral algebra, so that $\mathbb{V}(\mathcal{T})$ is simple when thought of as a module over itself.
Finally, a technical hypothesis that we make throughout this work, which holds in all known chiral algebras which come from four dimensions, is that $\mathbb{V}(\mathcal{T})$ is strongly finitely generated, which by definition means that there are only finitely many operators which do not appear in the non-singular part of any OPE.

The two-dimensional chiral algebra $\mathbb{V}(\mathcal{T})$ witnesses many features of the physics of its four-dimensional parent theory $\mathcal{T}$. For example, the entire Higgs branch $\mathcal{M}_H(\mathcal{T})$, which by definition is the subspace of the full moduli space of $\mathcal{T}$ on which $\mathfrak{su}(2)_R$ is broken, can conjecturally be recovered from $\mathbb{V}(\mathcal{T})$ as its associated variety~\cite{Beem:2017ooy}. We now define this notion. Our convention is that the modes of an operator in a VOA $\mathcal{V}$ are defined through
\begin{align}
    \varphi(z) = \sum_{n\in \mathbb{Z}} \varphi_n z^{-n-h}.
\end{align}
Then, one introduces the subspace 
\begin{align}
    C_2(\mathcal{V}) = \mathrm{span}\{ \varphi_{-h_{\varphi}-1}\chi \mid \varphi,\chi \in\mathcal{V}\} \subset \mathcal{V}.
\end{align}
The quotient $R_{\mathcal{V}}:=\mathcal{V}\big/ C_2(\mathcal{V})$ inherits the structure of a commutative, associative Poisson algebra, where the product is given by normal ordering, 
\begin{align}
    \mathrm{NO}(\varphi,\chi) = \varphi_{-h_{\varphi}}\chi_{-h_{\chi}}|0\rangle 
\end{align}
and the bracket is defined as 
\begin{align}
    \{\varphi,\chi\} = \varphi_{-h_{\varphi}+1}\chi.
\end{align}
We write $(R_{\mathcal{V}})_{\mathrm{red}}$ for the quotient of $R_{\mathcal{V}}$ by the ideal generated by its nilpotent elements. By definition, the associated variety of $\mathcal{V}$ is then~\cite{arakawa2012remark}
\begin{align}
    X_{\mathcal{V}} = \mathrm{Spec}(R_{\mathcal{V}})_{\mathrm{red}}.
\end{align}
\noindent \textbf{Conjecture (Higgs branch reconstruction)~\cite{Beem:2017ooy}:} The ring $(R_{\mathbb{V}(\mathcal{T})})_{\mathrm{red}}$ is the coordinate ring of the Higgs branch $\mathcal{M}_H(\mathcal{T})$ of $\mathcal{T}$. In other words, $\mathcal{M}_H(\mathcal{T}) \cong X_{\mathbb{V}(\mathcal{T})}.$\\

\noindent See \cite{Arakawa:2018egx} for a proof of Higgs branch reconstruction in the special case of genus zero class $\mathcal{S}$ theories. 

The significance of this conjecture for our analysis is that it immediately implies a useful finiteness condition on $\mathbb{V}(\mathcal{T})$. A VOA $\mathcal{V}$ is said to be quasi-lisse if its associated variety $X_{\mathcal{V}}$ has finitely many symplectic leaves; it is said to be lisse (or $C_2$-cofinite \cite{arakawa2012remark}, which is believed to be a necessary condition for rationality) if the associated variety $X_{\mathcal{V}}$ is a point. The Higgs branch of an SCFT will always have finitely many symplectic leaves, and so we learn that a VOA coming from four-dimensions is quasi-lisse if we assume the Higgs branch reconstruction conjecture. As a special case, an SCFT without a Higgs branch has a $C_2$-cofinite VOA. To avoid word salads, when we say that a VOA is quasi-lisse, we will implicitly assume that it is also simple, strongly finitely generated, and of CFT-type.

Quasi-lisse VOAs are well-behaved generalizations of rational VOAs. In particular, they retain some of the distinctive properties of rational VOAs which make them so tractable. Crucially for our purposes, Arakawa and Kawasetsu \cite{Arakawa:2016hkg} have shown that, for any  quasi-lisse chiral algebra $\mathcal{V}$ with integer conformal dimensions, there exists a finite-order monic modular differential equation (MDE) which annihilates the character of any ordinary  $\mathcal{V}$-module $M$ \footnote{An ordinary module is essentially one on which $L_0$ acts semi-simply with finite-dimensional eigenspaces and spectrum bounded from below. See e.g.\ \cite{abe2004rationality} for a more precise definition.}, including the character of $\mathcal{V}$ itself. (See e.g.\ \cite{Mathur:1988na} for early work on MDEs in 2d conformal field theory.)  That is, there exists an integer $n$ and holomorphic modular forms $g_\ell(\tau)$ of weight $2\ell$ such that 
\begin{align}\label{eqn:MDE}
    \left(\mathcal{D}^{(n)}+\sum_{\ell=1}^{n}g_\ell(\tau)\mathcal{D}^{(n-\ell)}\right) \mathrm{ch}_M(\tau) =0.
\end{align}
Here, $\mathrm{ch}_M(\tau)$ is the character of the ordinary $\mathcal{V}$-module $M$,
\begin{align}
    \mathrm{ch}_M(\tau) = \mathrm{Tr}_M(-1)^Fq^{L_0-c_{2\mathrm{d}}/24}
\end{align}
and 
\begin{align}
\begin{split}
    \mathcal{D}^{(\ell)} &= \partial_{(2\ell-2)} \circ \partial_{(2\ell-4)}\circ\cdots\circ \partial_{(0)} \\ 
    &\partial_{(\ell)}=q\frac{d}{dq}-\frac{\ell}{12}E_2(\tau)
\end{split}
\end{align}
where $E_2(\tau) = 1+\cdots$ is the normalized weight 2 Eisenstein series. In the case that $\mathcal{V}$ contains operators with half-integer conformal dimensions, a similar statement can be shown to be true \cite{Li:2023uot}, except one must take the $g_\ell(\tau)$ in Equation \eqref{eqn:MDE} to be holomorphic modular forms for the congruence subgroup 
\begin{align}
    \Gamma^0(2) = \left\{ \left(\begin{smallmatrix} a & b \\ c & d\end{smallmatrix} \right) \in \textsl{SL}_2(\mathbb{Z})\mid b \equiv 0 ~\mathrm{mod}~2 \right\}.
\end{align}

\subsection*{Rationality of $c$}

An important  implication of the MDE in Equation \eqref{eqn:MDE} is that the characters $\mathrm{ch}_M(\tau)$ of simple ordinary modules $M$ participate as the components of a finite-dimensional vector-valued modular form. In the context of two-dimensional rational conformal field theory, Anderson and Moore \cite{Anderson:1987ge}  demonstrated that it is essentially this fact which implies that the central charge and conformal dimensions are rational numbers.
One salient difference between rational VOAs and quasi-lisse VOAs is that the vector-valued modular form in which the characters $\mathrm{ch}_M(\tau)$ participate generally has components with logarithmic contributions, a fact which owes its origins to the non-semisimplicity of the matrix assigned to $T=\left(\begin{smallmatrix} 1 & 1 \\ 0 & 1\end{smallmatrix}\right)$ by the modular representation with respect to which the characters transform.  We will show that the argument of Anderson and Moore, with modest modifications, survives in the presence of these logarithmic contributions~\footnote{Miyamoto stated that the same conclusion applies to $C_2$-cofinite but not necessarily rational VOAs (Corollary 5.10 of \cite{miyamoto2004modular}), but details of the proof were omitted. We present here an argument in the more general quasi-lisse case.}.  

A complex number $x$ which lies outside of $\mathbb{Q}$ can be detected using algebraic automorphisms. An algebraic automorphism is a map $\phi:\mathbb{C}\to \mathbb{C}$ which preserves multiplication and addition, $\phi(x+y)=\phi(x)+\phi(y)$ and $\phi(xy)=\phi(x)\phi(y)$. It follows straight-forwardly from the definition that an algebraic automorphism fixes the rational numbers, i.e.\ $\phi(p/q)=p/q$ for all $p/q \in \mathbb{Q}$. In fact, the converse is true as well: if $x$ is any non-rational complex number, then there always exists an algebraic automorphism $\phi$ such that $\phi(x)\neq x$. 

Algebraic automorphisms can be used to detect non-rational exponents in the components of a vector-valued modular form as well. To proceed, we follow Anderson and Moore in calling a holomorphic function $f:\mathbb{H}\to\mathbb{C}$ quasi-automorphic if 
\begin{enumerate}
    \item $\mathrm{Span}\{f(\gamma\cdot \tau)\}_{\gamma\in\textsl{SL}_2(\mathbb{Z})}$ is finite-dimensional, and 
    \item $f$ satisfies a particular growth condition. Namely, for all real numbers $a<b$ and $C>0$, there exist real numbers $A,B>0$ such that 
    \begin{align}
        |f(\tau)|\leq Ae^{B\mathrm{Im}(\tau)}, \ \ \ a\leq \mathrm{Re}(\tau) \leq b, \ \ \ \mathrm{Im}(\tau)\geq C.
    \end{align}
\end{enumerate}
Note that any solution of a finite-order MDE  is quasi-automorphic, even when the coefficient functions $g_\ell(\tau)$ of the MDE are only modular with respect to the congruence subgroup $\Gamma^0(2)$. Furthermore, it is possible to show (Proposition 2 of \cite{Anderson:1987ge}) that if $f$ is quasi-automorphic and admits a non-logarithmic $q$-expansion,
\begin{align}
    f(\tau) = \sum_i s_i q^{\omega_i},
\end{align}
then its conjugate with respect to an algebraic automorphism $\phi$,
\begin{align}
    f^\phi(\tau) \equiv \sum_i \phi(s_i) q^{\phi(\omega_i)},
\end{align}
is also quasi-automorphic. 

We apply these considerations to the character $\mathrm{ch}_M(\tau)$ of an ordinary simple module $M$ of a  quasi-lisse VOA $\mathcal{V}$. Because it is a solution to an MDE, it is quasi-automorphic, and it moreover admits a non-logarithmic $q$-expansion of the shape 
\begin{align}
    \mathrm{ch}_M(\tau) = q^{-c_{2\mathrm{d}}/24+h_M}\sum_{n\in \mathbb{Z}_{\geq 0}}d_{n/2}q^{n/2}
\end{align}
where $h_M$ is the conformal dimension of $M$, and the coefficients $d_{n/2}$ are integers as they are dimensions of graded-components of $M$. Now, assume by way of contradiction that the combination $\theta\equiv -c_{2\mathrm{d}}/24+h_M$ is not a rational number. Then there exists some algebraic automorphism $\phi$ which is capable of detecting the irrationality of $\theta$, in the sense that $\phi(\theta)\neq \theta$. Conjugating the character of $M$ by this automorphism leads to the relation
\begin{align}\label{eqn:AMequation}
    \mathrm{ch}_M^\phi(\tau) =e^{\alpha\tau}\mathrm{ch}_M(\tau)
\end{align}
where $\alpha=2\pi i (\phi(\theta)-\theta) \neq 0$. On the other hand, performing a modular transformation on both sides of Equation \eqref{eqn:AMequation} recovers
\begin{align}\label{eqn:AMequation2}
    \tilde{\mathrm{ch}}_M^\phi(\tau) = e^{-\alpha/\tau} \tilde{\mathrm{ch}}_M(\tau)
\end{align}
where we have defined $\tilde{f}(\tau)\equiv f(-1/\tau)$ for any function $f:\mathbb{H}\to\mathbb{C}$.  Because both $\mathrm{ch}_M(\tau)$ and $\mathrm{ch}_M^\phi(\tau)$ are quasi-automorphic, it follows that their $\tau\to-1/\tau$ transformations are as well, and hence admit logarithmic $q$-expansions of the form
\begin{align}\label{eqn:logarithmicexpansion}
\begin{split}
    \tilde{\mathrm{ch}}_M(\tau) &= \sum_j  P_j(\tau) q^{z_j}, \\
    \tilde{\mathrm{ch}}^\phi_M(\tau)& = \sum_k Q_k(\tau) q^{w_k},
\end{split}
\end{align}
where the $P_j$ and $Q_k$ are polynomials whose degrees are bounded from above by some non-negative integer $\Delta$, and the $z_j\equiv x_j+iy_j$ and $w_k\equiv u_k+iv_k$ are complex numbers whose real parts are bounded from below. (See e.g.\ \cite{knopp2009logarithmic} for a review.) We would like to argue that these logarithmic $q$-expansions are incompatible with Equation \eqref{eqn:AMequation2}, unless $\alpha=0$ for all algebraic automorphisms $\phi$, in which case $\theta$ is rational.

Let $J=\{j \mid x_j\leq x_{j'}, ~\forall j'\}$ and let $j_0\in J$ be the unique index such that $y_{j_0}\leq y_{j}$ for all $j \in J$. Define $k_0$ similarly. We take $\tau = (\cos\varphi +i\sin\varphi)t$ with $\varphi\ll1$ and $t\gg 1$ both real numbers. In this limit, up to exponentially small corrections, Equation \eqref{eqn:AMequation2} reads (after dividing by $\tau^\Delta$) as
\begin{align}
    \tilde{P}_{j_0}(1/\tau) q^{z_{j_0}}+\cdots =e^{-\alpha/\tau} \tilde{Q}_{k_0}(1/\tau) q^{w_{k_0}}+\cdots 
\end{align}
where $\tilde{P}_{j_0}(1/\tau) \equiv P_{j_0}(\tau)/\tau^\Delta$, and similarly for $\tilde{Q}_{k_0}(1/\tau)$.
Consistency of this equation requires that $z_{j_0}=w_{k_0}$ and that $\tilde{P}_{j_0}(1/\tau) = e^{-\alpha/\tau}\tilde{Q}_{k_0}(1/\tau)$ when expanded to any finite order in $1/\tau$. However the second condition is impossible because $e^{-\alpha/ \tau}$ is not a rational function of $1/\tau$, and so we reach a contradiction.

Therefore, it must be the case that $\theta=-c_{\mathrm{2d}}/24+h_M$ is rational. Because the vacuum has conformal dimension $h_{\mathcal{V}}=0$, we find that $c_{\mathrm{4d}}=-c_{\mathrm{2d}}/12$  and the conformal dimensions $h_M$ of simple ordinary modules are all rational separately. This proof applies not just to VOAs which descend from four dimensions, but also to any  quasi-lisse VOA. \\

\noindent \textbf{Summary}: Any (simple, CFT-type, strongly finitely-generated) quasi-lisse  VOA provably has a rational central charge $c_{2\mathrm{d}}$, and rational conformal dimensions $h_M$ for its ordinary simple modules. Assuming the Higgs branch reconstruction conjecture, or more conservatively that the Schur index has a finite orbit under modular transformations, it follows that $c_{4\mathrm{d}}$ is rational in any 4d $\mathcal{N}=2$ SCFT as well.

\subsection*{Rationality of $a$}
Having established the rationality of $c_{4\mathrm{d}}$, we now turn to the $a_{4\mathrm{d}}$ central charge. Di Pietro and Komargodski \cite{DiPietro:2014bca} have argued using high-temperature effective field theory that the asymptotics of the Schur index are universally controlled by the anomalies of the theory, namely
\begin{align}\label{eqn:DPK}
    \lim_{\tau \to 0 }\log \mathcal{I}_{\mathrm{schur}}(\tau) \sim \frac{4\pi i(c_{\mathrm{4d}}-a_{\mathrm{4d}})}{\tau}.
\end{align}

On the other hand, standard Cardy-like arguments \cite{Cardy:1986ie} invoking modular covariance can be brought to bear as well. Indeed, the fact that $\mathcal{I}_{\mathrm{Schur}}(\tau)$ is annihilated by an MDE implies that there is a vector-valued modular form with components $f_i(\tau)$ and $f_0(\tau) = \mathcal{I}_{\mathrm{Schur}}(\tau)$ such that
\begin{align}
    f_i(-1/\tau) = \sum_j \mathcal{S}_{ij} f_j(\tau).
\end{align}
In particular, this implies that the high-temperature behavior is governed by the equation
\begin{align}\label{eqn:Cardy}
     \lim_{\tau \to 0 }\log \mathcal{I}_{\mathrm{schur}}(\tau) \sim \frac{\pi i (c_{2\mathrm{d}}-24 h_\ast)}{12\tau}
\end{align}
where $h_\ast$ is the smallest exponent among those arising in the $q$-expansions of the components $f_i(\tau)$ with $\mathcal{S}_{0i}\neq 0$. Comparing Equation \eqref{eqn:Cardy} to Equation \eqref{eqn:DPK}, one finds the relation 
\begin{align}\label{eqn:a4d}
   a_{\mathrm{4d}}=\frac{1}{48}(24h_\ast-5c_{\mathrm{2d}}).
\end{align}
Actually, while this equation is believed to be true when $c_{\mathrm{4d}}>a_{\mathrm{4d}}$, there are known counter-examples in the more general case \cite{ArabiArdehali:2015ybk}.
However, in Lagrangian theories one can show \cite{AMR} that Equation \eqref{eqn:a4d} always holds for some $h_\ast$ which is congruent to one of the exponents of the $f_i(\tau)$ modulo $1/2$. With this modification, it is obeyed in all known ${\cal N} =2$ SCFTs (Lagrangian and non), and we take it as a technical assumption from now on.

Now, one might be tempted to declare victory on the grounds that both $c_{2\mathrm{d}}$ and the conformal dimensions $h_M$ of simple ordinary modules $M$ were proved to be rational in the previous section. However, this is slightly too quick. While the existence of a monic MDE is sufficient to determine that the characters $\mathrm{ch}_M(\tau)$ participate as the components of a vector-valued modular form, it does not show that they are the \emph{only} components. Indeed, there are known examples \cite{Beem:2017ooy} where the solution space of an MDE which annihilates $\mathcal{I}_{\mathrm{Schur}}(\tau)$ is polluted by functions $f_i(\tau)$  with non-rational coefficients and exponents which are therefore unrelated to any ordinary simple module $M$. Conjecturally, these $f_i(\tau)$ are not in the modular orbit of the vacuum character $\mathcal{I}_{\mathrm{Schur}}(\tau)$, and so would not spoil the approach of using Equation \eqref{eqn:a4d} to deduce the rationality of $a_{\mathrm{4d}}$. However, to the best of our knowledge, this conjecture has yet to be established.

We can instead proceed as follows. Consider first a 4d $\mathcal{N}=2$ SCFT without a Higgs branch. In this case, it follows from the Higgs branch reconstruction conjecture that the protected chiral algebra is lisse/$C_2$-cofinite, and in this case the representation theory is under comparatively better control \cite{miyamoto2004modular} (see also \cite{Creutzig:2016fms} for a nice review). In particular, it is known that 
\begin{align}
    \mathrm{ch}_M(-1/\tau) = \sum_N P_{M,N}(\log q)\mathrm{ch}_N(\tau)
\end{align}
for any simple module $M$, where the $P_{M,N}(x)$ are polynomials in $x$. Applying this to $M=\mathbb{V}(\mathcal{T})$, one learns that the modular transformation of $\mathcal{I}_{\mathrm{Schur}}(\tau)$ involves only other characters of $\mathbb{V}(\mathcal{T})$, whose conformal dimensions are all rational \cite{miyamoto2004modular}. Therefore, the naive application of Equation \eqref{eqn:a4d} successfully yields the rationality of $a_{\mathrm{4d}}$.

We may generalize to a 4d $\mathcal{N}=2$ SCFT $\mathcal{T}_{\mathrm{UV}}$ with a non-empty Higgs branch as follows. Note that by going to a generic point on the Higgs branch, the theory flows in the IR to a (possibly trivial) SCFT $\mathcal{T}_{\mathrm{IR}}$ without a Higgs branch, along with $\dim_{\mathbb{H}}(\mathcal{M}_H(\mathcal{T}_{\mathrm{UV}}))$ decoupled free hyper multiplets. The $\mathfrak{u}(1)_r$ symmetry is unbroken on the Higgs branch, and so we may apply anomaly matching arguments to it. In particular, $\mathfrak{u}(1)_r$ possesses a cubic anomaly of the form
\begin{align}\label{eqn:U(1)rcubicanomaly}
    \partial_\mu J^\mu = \frac{3(a_{4\mathrm{d}}-c_{4\mathrm{d}})}{\pi^2} V_{\mu\nu} \tilde{V}^{\mu\nu},
\end{align}
where $J^\mu$ is the $\mathfrak{u}(1)_r$ current and $V_{\mu\nu}$ is the field strength of a background $\mathfrak{u}(1)_r$ gauge field. Therefore, $a_{4\mathrm{d}}-c_{4\mathrm{d}}$ is matched between the UV and the IR, leading to the relation
\begin{align}\label{eqn:anomalymatchinghiggs}
\begin{split}
   & (a_{4\mathrm{d}})_{\mathrm{UV}}=(c_{4\mathrm{d}})_{\mathrm{UV}} +\\
   & \hspace{.5in} (a_{4\mathrm{d}}-c_{4\mathrm{d}})_{\mathrm{IR}}-\frac{1}{24}\dim_{\mathbb{H}}(\mathcal{M}_H(\mathcal{T}_{\mathrm{UV}})), ,
\end{split}
\end{align}
where we have used the fact that $24(c_{4\mathrm{d}}-a_{4\mathrm{d}})=1$ for a free hypermultiplet.
The Weyl anomalies appearing on the right-hand side are all rational by the results of the previous section; The Euler anomaly $(a_{4\mathrm{d}})_{\mathrm{IR}}$ is that of a theory without a Higgs branch, and so is rational by the argument of the previous paragraph. 
Thus, we can conclude from Equation \eqref{eqn:anomalymatchinghiggs} that the Euler anomaly $(a_{4\mathrm{d}})_{\mathrm{UV}}$ of our UV SCFT is rational as desired. \\

\noindent \textbf{Summary}: As a technical hypothesis on the high-temperature limit of the Schur index, assume that Equation \eqref{eqn:a4d} is  true in any theory without a Higgs branch for some $h_\ast$ which is equal (modulo $1/2$) to the conformal dimension of a simple module of the protected chiral algebra. Then $a_{\mathrm{4d}}$ is rational in any 4d $\mathcal{N}=2$ SCFT, including theories with non-empty Higgs branches. \\

\subsection*{Future Direction: Rationality of $k$}

As mentioned in the introduction, the Rationality Conjecture is expected to also apply to the central charge $k_{4\mathrm{d}}=-2k_{2\mathrm{d}}$ of any simple non-Abelian flavor symmetry $\mathfrak{g}$ of a 4d $\mathcal{N}=2$ SCFT. The methods employed in this paper suggest two possible approaches to proving this.

The first approach involves studying the modular properties of the flavored Schur index, i.e.\ the vacuum character of the protected chiral algebra refined by fugacities with respect to the symmetry $\mathfrak{g}$. It is expected that in any quasi-lisse VOA with an affine Kac--Moody subalgebra $(\mathfrak{g})_{k_{2\mathrm{d}}}$, the flavored vacuum character participates as a component of a logarithmic vector-valued  Jacobi form, with index which is sensitive to the level $k_{2\mathrm{d}}$. (See \cite{Krauel:2013lra} for a theory of flavored characters in the setting of strongly rational VOAs). One may attempt to place this expectation on firmer mathematical footing, perhaps by arguing that the vacuum character always satisfies a flavored modular differential equation using methods similar to those of \cite{Arakawa:2016hkg} (see \cite{Pan:2021ulr,Zheng:2022zkm,Pan:2023jjw} for recent studies of flavored MDEs in the context of 4d $\mathcal{N}=2$ SCFTs). Just as the modularity of the unflavored Schur index was sufficient to establish the rationality of $c_{4\mathrm{d}}$, one might hope that the transformation properties of the flavored Schur index likewise imply the rationality of $k_{4\mathrm{d}}$.

A second approach involves studying anomaly matching on the Higgs branch. Any 4d $\mathcal{N}=2$ SCFT with a continuous global symmetry admits moment map operators, which are the Higgs branch operators in the same supermultiplet as the conserved currents. In situations where one is able to consistently activate VeVs for just the moment maps without turning on other moduli (the case of ``nilpotent Higgsing''), the flavor central charge $k_{4\mathrm{d}}$ arises in various mixed anomalies between the R-symmetries and flavor symmetries; anomaly matching then implicates $k_{4\mathrm{d}}$ in an equation involving the UV and IR conformal central charges, which we know to be rational~\footnote{Nilpotent Higgsing has a direct 2d reinterpretation as quantum Drinfeld-Sokolov reduction for modules~\cite{Beem:2014rza}. The anomaly-matching formula for central charges is reproduced by rigorous VOA considerations.}. It is however not clear that such a pattern of Higgsing is universally available (though we are not aware of a clear cut counterexample). One would need to either establish universality of nilpotent Higgsing or study anomaly matching in the general geometric situation.   \\

\noindent \textbf{Acknowledgments} We gratefully acknowledge discussions with Tomoyuki Arakawa, Arash Ardehali, Christopher Beem, Anirudh Deb, Mykola Dedushenko, and Zohar Komargodski. We also thank Arash Ardehali for helpful comments on a draft.

\bibliographystyle{plainnat}
\bibliography{main}

\end{document}